\documentclass[showpacs,preprintnumbers,amsmath,amssymb,eqsecnum]{revtex4}

\newcommand{\beq}{\begin{equation}}
\newcommand{\beqa}{\begin{eqnarray}}
\newcommand{\eeq}{\end{equation}}
\newcommand{\eeqa}{\end{eqnarray}}
\newcommand{\simg}{\gtrsim}

\newenvironment{namelist}[1]{%
 \begin{list}{}
  {
  \settowidth{\labelwidth}{#1}
  \setlength{\leftmargin}{1.1\labelwidth}}
}{%
 \end{list}}

\usepackage{graphicx}
\usepackage{dcolumn}
\usepackage{bm}

\begin{document}

\preprint{KUNS-, OU-TAP , YITP-}

\title{
Grad-Shafranov equation in noncircular stationary axisymmetric spacetimes
}

\author{Kunihito Ioka$^{1}$}
\author{Misao Sasaki$^{1}$}
\affiliation{%
$^{1}$Department of Earth and Space Science, Osaka 
University, Toyonaka 560-0043, Japan}

\date{\today}

\begin{abstract}
A formulation is developed for general relativistic ideal 
magnetohydrodynamics in stationary axisymmetric spacetimes.
We reduce basic equations to a single second-order
partial differential equation, the so-called Grad-Shafranov (GS) equation.
Our formulation is most general in the sense that
it is applicable even when a stationary axisymmetric
spacetime is noncircular,
that is, even when it is impossible to foliate a spacetime
with two orthogonal families of two-surfaces.
The GS equation for noncircular spacetimes is crucial for the 
study of relativistic stars with a toroidal magnetic
field or meridional flow, such as magnetars,
since the existence of a toroidal field or meridional flow violates
the circularity of a spacetime.
We also derive the wind equation in noncircular spacetimes,
and discuss various limits of the GS equation.
\end{abstract}

\maketitle

\section{Introduction}\label{sec:intro}
While most neutron stars have magnetic fields of $\sim 10^{12}$-$10^{13}$ G,
studies of soft gamma-ray repeaters and anomalous X-ray pulsars
suggest that a significant fraction $(\simg 10 \%)$ of neutron stars
is born with larger magnetic fields $\sim 10^{14}$-$10^{15}$ G
\cite{k98,dt92,m02,t01}.
The internal magnetic field of a new born neutron star
may be even larger $\simg 10^{16}$ G
if it is generated by the helical dynamo \cite{dt92,td93}.
A magnetic field of nearly maximum strength
allowed by the virial theorem $\sim 10^{17}$ G
may be achieved if the central engine of gamma-ray bursts 
are neutron stars \cite{u92,n98,kr98,wyhw00}.
In such super-magnetized neutron stars, so-called magnetars
\cite{dt92,t01},
the magnetic fields have substantial effects 
on their internal stellar structure.
Especially the deformation due to the magnetic stress 
becomes non-negligible \cite{b95,bg96,i01,kok99}.
Since the deformation affects the precession, oscillations
and the gravitational wave emission of neutron stars
\cite{i01,bg96,m99,c02,it00}, it is important to investigate
equilibrium configurations of magnetars.

General relativistic effects are sizable in the interior of 
a neutron star, so that any quantitative investigation of the magnetars
has to be based on general relativistic magnetohydrodynamics (MHD)
\cite{l67,nt73,bo78,bo79}.
Therefore, we have to solve the matter and electromagnetic field
configurations in a curved spacetime, 
and have to take account of the electromagnetic energy-momentum 
as a source of the gravitational field.
So far several works have been devoted to equilibrium configurations
of a magnetized star in a stationary axisymmetric spacetime
\cite{b93,b95,kok99,cpl01}.
However these works consider only poloidal magnetic fields for simplicity,
since the existence of only a poloidal field is compatible with the
circularity of the spacetime \cite{gb93,o02}.
In a circular spacetime, there exists a family of two-surfaces everywhere
orthogonal to the plane defined by the two Killing vectors
associated with stationarity $\eta^\mu=(\partial/\partial t)^\mu$
and axisymmetry $\xi^\mu=(\partial/\partial \varphi)^\mu$
 \cite{p66,c69,c72}.
Thus one may choose the coordinates 
$\left(x^{\mu}\right)=\left(t,x^1,x^2,\varphi\right)$
such that the metric components $g_{01}$, $g_{02}$, $g_{31}$
and  $g_{32}$ are identically zero.
As a consequence, the problem is simplified dramatically.

However non-negligible toroidal magnetic fields are likely to exist
in nature.
Differential rotation generated during the gravitational 
collapse \cite{s03,s00} or in the binary coalescence \cite{su00} may
wind up the frozen-in magnetic field to amplify the toroidal component
\cite{kr98,wyhw00}. A toroidal magnetic field may be generated
by the $\alpha$-$\Omega$ dynamo during the first few seconds
after the formation of a millisecond pulsar \cite{dt92,td93}.
In addition, convective motion may also exist in the interior of a
neutron star \cite{meas76},
which also violates the circularity of the spacetime \cite{gb93,o02}.
Thus, we have to consider noncircular spacetimes.
The degree of noncircularity of the spacetime
in a neutron star with mass $M_{*}$ and radius $R_{*}$,
will be about $\sim (M_{*}/R_{*}) v_{\rm mf}$
and $\sim 0.1$-$0.01 (M_{*}/R_{*}) {\cal R}_{M}$,
where $v_{\rm mf}$ is the velocity of the meridional flow
and ${\cal R}_{M}$ is the ratio of the magnetic energy
to the gravitational energy \cite{Ioka2}.

The problem to obtain an equilibrium configuration of a magnetar
can be separated into two parts.
The first part is the Einstein equations which determine the
spacetime geometry under a given configuration of matter and
electromagnetic fields.
The second part is the matter and electromagnetic field equations
in a given spacetime geometry.
A $(2+1)+1$ formalism to solve the Einstein equations under the presence
of a spatial Killing vector was developed by Maeda et al.~\cite{m80,nok87}
and by Sasaki~\cite{s84}.
This formalism is similar to the well-known $3+1$ formalism \cite{sy78}.
 Later Gourgoulhon and Bonazzola~\cite{gb93} developed a similar but 
different $(2+1)+1$ formalism which is more suited for stationary axisymmetric
spacetimes. So here we focus our attention on the second problem, i.e., 
to formulate the equations of motion of matter and electromagnetic fields
in a curved spacetime.

It is well known that the basic equations for a
stationary axisymmetric ideal MHD system can be
reduced to a single second-order, nonlinear partial differential equation,
the so-called Grad-Shafranov (GS) equation, 
for a quantity called the flux function, $\Psi$.
The GS equation was derived in the Newtonian case \cite{l86}, 
the Schwarzschild spacetime case \cite{ml86},
and the Kerr spacetime case \cite{n91,b97}.
The flux function $\Psi$ is such that it is constant over each surface
generated by rotating the magnetic field lines (or equivalently the flow lines)
about the axis of symmetry and 
the GS equation determines the transfield equilibrium.
Any physical quantities can be calculated from the solution $\Psi$
of the GS equation. However, the GS equation in noncircular spacetimes 
has never been derived explicitly.

In this paper we derive the GS equation explicitly in noncircular
(i.e., the most general) stationary axisymmetric spacetimes.
This is a first step toward the study of equilibrium configurations
of magnetars. An attempt to solve the GS equation will be discussed 
in a subsequent paper. This paper is organized as follows.
In Sec.~\ref{sec:basic}, we briefly review the conservation laws
in stationary axisymmetric general relativistic ideal MHD systems
\cite{bo78,bo79} that are used to characterize
the matter and electromagnetic field configurations.
We neglect dissipative effects, which is a reasonable assumption
because of the high conductivity and the low viscosity in neutron stars.
In Sec.~\ref{sec:gscom}, we derive the GS equation in an un-elucidated
form. At this stage it is not clear if the GS equation is a second-order
differential equation for the flux function $\Psi$.
In Sec.~\ref{sec:2+1+1},
we briefly review the $(2+1)+1$ formalism by
Gourgoulhon and Bonazzola~\cite{gb93} to describe the geometry
of noncircular stationary axisymmetric spacetimes in a 
transparent way. We do not, however, discuss the Einstein 
equations but assume the geometry to be given.
In Sec.~\ref{sec:inv}, we explicitly demonstrate that all physical
quantities except for the metric can be evaluated from the 
flux function $\Psi$.
In Sec.~\ref{sec:GScov}, we write down the GS equation
in the covariant form projected onto the 2-surface $\Sigma_{t\varphi}$
spanned by $t=$const. and $\varphi=$const..
We also discuss various limits of the GS equation.
Finally, we summarize our result in Sec.~\ref{sec:sum}.
The energy-momentum tensor decomposed in the $(2+1)+1$ form
is given in Appendix~\ref{sec:emt2+1+1}, and
notation and symbols are summarized in 
Appendix~\ref{sec:symbols}.

We use the units $c=G=k_B=1$.
Greek indices ($\mu,\,\nu,\,\alpha,\,\beta,\,\cdots$) run from $0$ to 
$3$, small Latin indices ($i,\,j,\,k,\,\cdots$) from $1$ to $3$,
and capital Latin indices ($A,\,B,\,C,\,\cdots$) from $1$ to $2$,
where $x^0=t$ and $x^3=\varphi$.
The signature of the 4-metric is $(-,+,+,+)$.

\section{Basic equations and conservation laws}
\label{sec:basic}

\subsection{Basic equations for general relativistic magnetohydrodynamics}

The basic equations governing a general relativistic ideal MHD system
are as follows \cite{l67,nt73}.
Baryons are conserved,
\beqa
(\rho u^{\mu})_{;\mu}=u^{\mu} \rho_{,\mu} + \rho {u^{\mu}}_{;\mu}=0,
\label{eq:bacon}
\eeqa
where $\rho$ is the rest mass density 
(i.e., the baryon mass times the baryon number density) and
$u^{\mu}$ is the fluid 4-velocity with 
\beqa
u_{\mu} u^{\mu}=-1.
\label{eq:norm}
\eeqa
The electromagnetic field is governed by the Maxwell equations,
\beqa
F_{[\mu \nu; \alpha]}&=&0,
\label{eq:maxwell1}
\\
{F^{\mu \nu}}_{;\nu}&=&4\pi J^{\mu},
\label{eq:maxwell2}
\eeqa
where $F_{\mu \nu}$ and $J^{\mu}$ are
the field strength tensor and the electric current 4-vector, respectively.
Equation~(\ref{eq:maxwell1}) implies the existence of
the vector potential $A_\mu$,
\beqa
F_{\mu \nu}=A_{\nu, \mu}-A_{\mu, \nu}.
\label{eq:vp}
\eeqa
The electric and magnetic fields in the fluid rest frame are defined
as
\beqa
E_{\mu}&=&F_{\mu \nu} u^{\nu},
\label{eq:E}
\\
B^{\mu}&=&-\frac{1}{2} \epsilon^{\mu \nu \alpha \beta}
u_{\nu} F_{\alpha \beta},
\label{eq:B}
\eeqa
where $\epsilon_{\mu \nu \alpha \beta}$ is the Levi-Civita 
antisymmetric tensor with $\epsilon_{0123}=\sqrt{-g}$.
Equations (\ref{eq:E}) and (\ref{eq:B}) are inverted to give
\beqa
F_{\mu \nu}=u_{\mu} E_{\nu}-u_{\nu} E_{\mu}
+\epsilon_{\mu \nu \alpha \beta} u^{\alpha} B^{\beta},
\label{eq:FuB}
\eeqa
with $E_{\mu} u^{\mu}=B_{\mu} u^{\mu}=0$.
In the ideal MHD, we assume the perfect conductivity, so that
\beqa
E_{\mu}&=&F_{\mu \nu} u^{\nu}=0.
\label{eq:ideal}
\eeqa
The equations of motion for the fluid are given by ${T^{\mu \nu}}_{;\nu}=0$,
where $T^{\mu \nu}$ is the total energy-momentum tensor
of the fluid and electromagnetic fields,
\beqa
T^{\mu \nu}=(\rho+\rho\epsilon+p) u^{\mu} u^{\nu}+p g^{\mu \nu}
+\frac{1}{4\pi} \left(F^{\mu \alpha} {F^{\nu}}_{\alpha}
-\frac{1}{4} g^{\mu \nu} F^{\alpha \beta} F_{\alpha \beta}\right).
\label{eq:emten}
\eeqa
Then we obtain the conservation of the fluid energy,
\beqa
u^{\mu} (\rho+\rho\epsilon)_{,\mu}+(\rho+\rho\epsilon+p) {u^{\mu}}_{;\mu}=0,
\label{eq:encon}
\eeqa
and the Euler equations,
\beqa
(\rho+\rho\epsilon+p) {u^{\mu}}_{;\nu} u^{\nu}
+(g^{\mu \nu}+u^{\mu} u^{\nu}) p_{,\nu}
-F^{\mu \nu} J_{\nu}=0,
\label{eq:euler}
\eeqa
where $\epsilon$ and $p$ 
are the internal energy per unit mass and pressure, respectively.
Eliminating ${u^{\mu}}_{;\mu}$ from Eqs.~(\ref{eq:bacon}) and 
(\ref{eq:encon}) gives
\beqa
u^{\mu} (\rho+\rho\epsilon)_{,\mu}=\mu u^{\mu} \rho_{,\mu},
\label{eq:adia}
\eeqa
where
\beqa
\mu=1+\epsilon+\frac{p}{\rho}\,,
\label{eq:defmu}
\eeqa
is the enthalpy per unit mass.
Assuming local thermodynamic equilibrium,
the first law of thermodynamics is given by
\beqa
d\epsilon=-p\,d\left(\frac{1}{\rho}\right) + T dS,
\label{eq:1law}
\eeqa
where $S$ and $T$ are the entropy per unit mass and the temperature.
Then Eqs.~(\ref{eq:defmu}) and (\ref{eq:1law}) imply
\begin{eqnarray}
d\mu=\frac{dp}{\rho}+TdS\,.
\label{eq:dmu}
\end{eqnarray}
Finally we supply the equation of state,
\beqa
p=p(\rho, S).
\label{eq:eos}
\eeqa

\subsection{Conservation laws in a stationary axisymmetric spacetime}

Here we recapitulate the conservation laws derived from
the basic equations in the previous subsection
in a stationary axisymmetric spacetime.
There exists two Killing vectors associated with stationarity 
and axisymmetry, which we denote by $\eta^\mu$ and $\xi^\mu$, respectively. 
The Lie derivatives of all physical quantities along the Killing vectors
must vanish, e.g., 
${\cal L}_{\xi}u^{\mu}=\xi^{\nu}{u^{\mu}}_{;\nu}-u^{\nu}{\xi^{\mu}}_{;\nu}=0$.
We take $\eta^\mu=(\partial/\partial t)^\mu$
and $\xi^\mu=(\partial/\partial\varphi)^\mu$ so that
$x^{0}=t$ and $x^{3}=\varphi$ are the time and azimuthal coordinates
associated with the Killing vectors $\eta^{\mu}$ and $\xi^{\mu}$,
respectively. Thus all physical quantities
are independent of $t$ and $\varphi$.

Bekenstein and Oron \cite{bo78,bo79} showed that a stationary
axisymmetric system has several conserved quantities along
each flow line. This is a general relativistic generalization of 
Ferraro's integrability condition \cite{fe37,fe54,wo59}.
By exploiting the gauge freedom to make $A_{\mu,\nu} \eta^{\nu}=A_{\mu,0}=0$ 
and $A_{\mu,\nu} \xi^{\nu}=A_{\mu,3}=0$,
we can show that the magnetic potential
$\Psi:=A_{\mu} \xi^{\mu}=A_{3}$ as well as the electric potential
$\Phi:=A_{\mu} \eta^{\mu}=A_{0}$ are constant along each flow line,
i.e., $u^{\mu} (\xi^{\nu} A_{\nu})_{,\mu}=
u^{\mu} (\eta^{\nu} A_{\nu})_{,\mu}=0$
\cite{be85}.
Henceforth we label the flow line by $\Psi$, which we refer
to as the flux function as in the non-relativistic case \cite{l86}.
The $\Psi={\rm const.}$ surfaces are called the flux surfaces,
which are generated by rotating the magnetic field lines 
(or the flow lines) about the axis of symmetry.

According to Bekenstein and Oron \cite{bo78,bo79},
one can show that
\beqa
F_{03}&=&0,
\label{eq:F03}
\\
F_{0A}&=&\Omega F_{A3},
\label{eq:F0A}
\\
F_{31}&=&-\Psi_{,1}=C\sqrt{-g} \rho u^{2},
\label{eq:F31}
\\
F_{23}&=&\Psi_{,2}=C\sqrt{-g} \rho u^{1},
\label{eq:F23}
\\
F_{12}&=&C\sqrt{-g} \rho (u^{3}-\Omega u^{0}),
\label{eq:F12}
\eeqa
where $\Omega(\Psi)$ and $C(\Psi)$ are conserved along each flow line
and hence are functions of the flux function $\Psi$.
The above equations are effectively first integrals 
of the Maxwell equations (\ref{eq:maxwell1}).
It may be useful to rewrite the above equations as
\beqa
B^{\mu}=-C \rho \left[(u_0 + \Omega u_3) u^{\mu} 
+ \eta^{\mu} + \Omega \xi^{\mu} \right].
\label{eq:Bu}
\eeqa
Note that $\Omega(\Psi)$ is the $\Psi$-derivative of the electric potential, 
$\Omega(\Psi)=-d \Phi/d \Psi$,
and coincides with the angular velocity
$d\varphi/dt=u^3/u^0=\Omega$ if there is no toroidal field $F_{12}=0$
and $C\ne 0$, from Eq.~(\ref{eq:F12}).
In addition, one can show that $E(\Psi)$, $L(\Psi)$ and $D(\Psi)$
are also conserved along each flow line \cite{bo78}, where
\beqa
-D&=&\mu (u_0 + \Omega u_3),
\label{eq:defD}
\\
-E&=&\left(\mu+\frac{B^2}{4\pi \rho}\right)
u_0 + C(u_0 + \Omega u_3) \frac{B_0}{4\pi},
\label{eq:defE}
\\
L&=&\left(\mu+\frac{B^2}{4\pi \rho}\right)
u_3 + C(u_0 + \Omega u_3) \frac{B_3}{4\pi},
\label{eq:defL}
\eeqa
and $B^2:=B_{\mu} B^{\mu}$.
Equations~(\ref{eq:defD}), (\ref{eq:defE}) and (\ref{eq:defL}) 
are first integrals of the equations of motion $T^{\mu\nu}{}_{;\nu}=0$
along $B_{\mu}$, $\eta_{\mu}$ and $\xi_{\mu}$, respectively.
In particular, Eq.~(\ref{eq:defE}) is 
a generalized Bernouilli's equation.
These conserved quantities are not mutually independent but
there is a relation among them \cite{bo78},
\beqa
D=E-\Omega L\,.
\label{eq:DEL}
\eeqa
Together with Eqs.~(\ref{eq:defD}) -- (\ref{eq:defL}),
this implies that
\beqa
\frac{B^2}{\rho}+C(B_0+\Omega B_3)=0.
\label{eq:B2}
\eeqa
With Eq.~(\ref{eq:B2}), we can rewrite $E(\Psi)$ and $L(\Psi)$ 
in Eqs.~(\ref{eq:defE}) and (\ref{eq:defL}) as
\beqa
-E&=&\mu u_{0}-\frac{1}{4\pi} C \Omega (u_{0} B_{3}-u_{3} B_{0}),
\label{eq:defE2}
\\
L&=&\mu u_{3}+\frac{1}{4\pi} C (u_{0} B_{3}-u_{3} B_{0}).
\label{eq:defL2}
\eeqa
Finally, from Eqs.~(\ref{eq:adia}) -- (\ref{eq:1law})
one finds that the entropy per unit mass $S$ is conserved along
each flow line, $u^{\mu} S_{,\mu}=0$, as a result of 
the perfect fluid form of the energy-momentum tensor. 
For a stationary axisymmetric spacetime, this implies that
$S$ is a function of $\Psi$, $S=S(\Psi)$.

In summary, for a give flux function $\Psi$,
there exists five conserved quantities, $E(\Psi)$, $L(\Psi)$, 
$\Omega(\Psi)$, $C(\Psi)$ and $S(\Psi)$.
Except for $S(\Psi)$, there are no perfectly relevant physical
interpretations of these quantities.
Nevertheless, by considering several limiting cases,
we may associate them with terms that describe their qualitative
nature. We may call $E(\Psi)$ the total energy per unit mass,
$L(\Psi)$ the total angular momentum per unit mass,
$\Omega(\Psi)$ the angular velocity,
and $C(\Psi)$ the magnetic field strength relative
to the magnitude of meridional flow.
 Since these conserved quantities are essentially
the first integrals of the equations of motion, specification of
these functions characterizes the configuration of the fluid flow 
and the electromagnetic field.
As we will see in Sec.~\ref{sec:inv}, all physical quantities are
completely determined once these conserved quantities are given
as functions of the flux function $\Psi$, provided that the spatial
configuration of $\Psi$ is known.
Therefore the problem reduces to solving an equation for the 
flux function that determines the spatial configuration of $\Psi$,
that is, the GS equation.

\section{Grad-Shafranov equation in the component expression}
\label{sec:gscom}

The GS equation is given by the transfield component
of the Euler equations~(\ref{eq:euler}).
In this section, we consider the $x^A$-derivative of the 
flux function in the Euler equations, and factorize the resulting
equation to derive the GS equation.
We express equations in terms of
their explicit coordinate components, since it is the most
straightforward way to incorporate the symmetry,
e.g., $(\cdots)_{,0}=(\cdots)_{,3}=0$.
Accordingly, the GS equation is given in the component
expression. A covariant form of the GS equation based on
$(2+1)+1$ formalism, which 
may be useful for numerical calculations, will be given 
in Sec.~\ref{sec:GScov}.

Using Eq.~(\ref{eq:dmu}), 
the Euler equations~(\ref{eq:euler}) can be written as
\beqa
\rho \mu u_{\mu;\nu} u^{\nu}+\rho\mu_{,\mu}
+\rho u_{\mu} u^{\nu} \mu_{,\nu}
-F_{\mu \nu} J^{\nu} - \rho T S_{,\mu}=0,
\label{eq:euler2}
\eeqa
where we have used that $u^{\mu} S_{,\mu}=0$.
First, let us consider the $x^1$-component of 
the Euler equations~(\ref{eq:euler2}).
The first term can be expressed as
\beqa
\rho \mu u_{1;\mu} u^{\mu}=
\rho \mu(u^1 u_{1;1}+u^{2} u_{1;2})
-\rho \mu \left(u^{0} \Gamma_{10}^{\mu} u_{\mu}
+u^{3} \Gamma_{13}^{\mu} u_{\mu}\right)\,,
\label{eq:1term}
\eeqa
where $\Gamma^{\alpha}_{\mu \nu}$ 
is the Christoffel symbol.
With Eq.~(\ref{eq:norm}), the third term in Eq.~(\ref{eq:euler2})
can be transformed as
\beqa
\rho u_{1} u^{\mu} \mu_{,\mu}&=&\rho u_{1} u^{1} \mu_{,1}+
\rho u_{1} u^{2} \mu_{,2}
\nonumber\\
&=&-\rho \mu_{,1}(1+u_{0} u^{0}+u_{2} u^{2}+u_{3} u^{3})
+\rho u_{1} u^{2} \mu_{,2}
\nonumber\\
&=&\rho u^{2}(u_{1} \mu_{,2}-u_{2} \mu_{,1})-\rho \mu_{,1}
-\rho u^{0} (\mu u_{0})_{,1}-\rho u^{3} (\mu u_{3})_{,1}
\nonumber\\
&&+\rho \mu u^{0}(u_{0;1}+\Gamma_{01}^{\mu} u_{\mu})
+\rho \mu u^{3}(u_{3;1}+\Gamma_{31}^{\mu} u_{\mu})
\nonumber\\
&=&\rho u^2\left[(\mu u_1)_{;2}-(\mu u_2)_{;1})\right]
-\rho\mu_{,1}
-\rho u^{0} (\mu u_{0})_{,1}-\rho u^{3} (\mu u_{3})_{,1}
\nonumber\\
&&-\rho\mu (u^1u_{1;1}+u^2u_{1;2})
+\rho\mu (u^0\Gamma_{01}^\mu u_\mu+u^3\Gamma_{13}^\mu u_\mu),
\label{eq:3term}
\eeqa
where we have used the fact $u^\mu u_{\mu;\nu}=0$.
Thus putting Eqs.~(\ref{eq:1term}) and (\ref{eq:3term}) together,
and using Eq.~(\ref{eq:F31}), we find
\beqa
\rho (\mu u_{1;\mu} u^{\mu} + \mu_{,1} + u_1 u^{\mu} \mu_{,\mu})
&=&\rho u^2 \left[(\mu u_{1})_{;2} - (\mu u_{2})_{;1}\right]
-\rho u^0 (\mu u_0)_{,1} - \rho u^3 (\mu u_3)_{,1},
\nonumber\\
&=&-\frac{1}{C\sqrt{-g}}
\left[(\mu u_{1})_{;2} - (\mu u_{2})_{;1}\right] \Psi_{,1}
-\rho u^0 (\mu u_0)_{,1} - \rho u^3 (\mu u_3)_{,1}\,.
\label{eq:euler1}
\eeqa
Next, the fourth term in Eq.~(\ref{eq:euler2}) can be transformed as
\beqa
-F_{1 \mu} J^{\mu}
&=&-F_{10}J^0-F_{13}J^3-\frac{1}{4\pi \sqrt{-g}} F_{1 A} (\sqrt{-g} F^{AB})_{,B}
\nonumber\\
&=&\left(\Omega J^0 - J^3\right) \Psi_{,1}
+\frac{1}{4\pi \sqrt{-g}} F_{12} (\sqrt{-g} F^{12})_{,1},
\label{eq:FJ}
\eeqa
where the second line follows from Eqs.~(\ref{eq:F0A}) and (\ref{eq:F31}).
{}From Eqs.~(\ref{eq:FuB}) and (\ref{eq:ideal}), we have
$F^{12}=-(u_0 B_3 - u_3 B_0)/\sqrt{-g}$.
Therefore, together with Eq.~(\ref{eq:F12}) that gives $F_{12}$,
the last term in Eq.~(\ref{eq:FJ}) is expressed as
\beqa
\frac{1}{4\pi \sqrt{-g}} F_{12} (\sqrt{-g} F^{12})_{,1}
&=&-\frac{1}{4\pi} C \rho (u^3 - \Omega u^0)(u_0 B_3 - u_3 B_0)_{,1}
\nonumber\\
&=&-\frac{1}{4\pi} \rho u^0 (u_0 B_3 - u_3 B_0) (C \Omega)' \Psi_{,1}
+\frac{1}{4\pi} \rho u^3 (u_0 B_3 - u_3 B_0) C' \Psi_{,1},
\nonumber\\
&-&\rho u^0 \left[-\frac{1}{4\pi} C \Omega (u_0 B_3 - u_3 B_0)\right]_{,1}
-\rho u^3 \left[\frac{1}{4\pi} C (u_0 B_3 - u_3 B_0)\right]_{,1},
\label{eq:f12f12}
\eeqa
where primes denote differentiation with respect to $\Psi$.
Finally, the last term in Eq.~(\ref{eq:euler2}) gives
\beqa
-\rho T S_{,1}=-\rho T S' \Psi_{,1}.
\label{eq:nTS}
\eeqa

Combining Eqs.~(\ref{eq:euler1}) -- (\ref{eq:nTS}),
we have
\beqa
&&\Biggl\{-\frac{1}{C\sqrt{-g}}
\left[(\mu u_{1})_{,2}-(\mu u_{2})_{,1}\right]
-\left(J^3-\Omega J^0\right)
\nonumber\\
&&-\frac{1}{4\pi} \rho u^{0} \left(u_{0} B_{3}-u_{3} B_{0}\right)
\left(C \Omega\right)'
+\frac{1}{4\pi} \rho u^{3} \left(u_{0} B_{3}-u_{3} B_{0}\right) C'
-\rho T S'\Biggr\} \Psi_{,1}
\nonumber\\
&&-\rho u^{0}\left[\mu u_{0}-\frac{1}{4\pi} C\Omega
\left(u_{0}B_{3}-u_{3}B_{0}\right)\right]_{,1}
-\rho u^{3} \left[\mu u_{3}+\frac{1}{4\pi} C
\left(u_{0}B_{3}-u_{3}B_{0}\right)\right]_{,1}
\label{eq:GStemp1}
=0.
\eeqa
Recalling the expressions for $E$ and $L$ given by Eqs.~(\ref{eq:defE2})
 and (\ref{eq:defL2}), respectively, we see that the last two terms 
are just their derivatives. Therefore, we can factor out
the $x^1$-derivative of the flux function in Eq.~(\ref{eq:GStemp1})
to obtain
\beqa
&&\Biggl\{-\frac{1}{C\sqrt{-g}}
\left[(\mu u_{1})_{,2}-(\mu u_{2})_{,1}\right]
-\left(J^3-\Omega J^0\right)
\nonumber\\
&&+\rho u^{0} \left[E'-\frac{1}{4\pi} \left(u_{0} B_{3}-u_{3} B_{0}\right)
\left(C \Omega\right)'\right]
-\rho u^{3}\left[L'-\frac{1}{4\pi} \left(u_{0} B_{3}-u_{3} B_{0}\right)
C'\right]-\rho T S'\Biggr\} \Psi_{,1}=0.
\label{eq:GSpsi}
\eeqa
The same analysis applies to the $x^2$-component of Eq.~(\ref{eq:euler2}),
and one finds the above equation~(\ref{eq:GSpsi}) 
with the replacement of $\Psi_{,1}$ by $\Psi_{,2}$.
Therefore, by assuming $\Psi_{,A}\ne 0$ ($A=1,2$),
the GS equation is given by
\beqa
J^3-\Omega J^0
+\frac{1}{C\sqrt{-g}}
\left[(\mu u_{1})_{,2}-(\mu u_{2})_{,1}\right]
-\rho u^{0} \left[E'-\Lambda\left(C \Omega\right)'\right]
+\rho u^{3}\left[L'-\Lambda C'\right]+\rho T S'=0\,,
\label{eq:GScom}
\eeqa
where, for convenience,
we have introduced an auxiliary quantity $\Lambda$ defined by
\begin{eqnarray}
\Lambda=\frac{1}{4\pi} (u_0 B_3 - u_3 B_0)\,.
\label{eq:Lambdadef}
\end{eqnarray}
At this stage, however, it is not clear if Eq.~(\ref{eq:GScom})
gives a second-order, nonlinear partial differential equation for 
the flux function $\Psi$, since the dependence on the 
flux function is unknown.
In Sec.~\ref{sec:inv}, we explicitly demonstrate that all the 
physical quantities appearing in the above equation
can be expressed in terms of $\Psi$ and its $x^A$-derivatives, 
and in Sec.~\ref{sec:GScov}
 we derive the GS equation in the covariant form
and we make it explicit that it is indeed 
a second-order, non-linear differential equation for $\Psi$.

\section{$(2+1)+1$ DECOMPOSITION}
\label{sec:2+1+1}
In this section, we briefly review the $(2+1)+1$ formalism
of the Einstein equations for stationary axisymmetric spacetimes
developed by Gourgoulhon and Bonazzola \cite{gb93},
in order to describe our metric in a covariant fashion.
Note that this formalism is different from the $(2+1)+1$ formalism
by Maeda, Sasaki, Nakamura, and Miyama \cite{m80,nok87} and Sasaki \cite{s84},
which is suitable to the axisymmetric gravitational collapse.
Here we adopt the formalism by Gourgoulhon and Bonazzola because 
it is more convenient for a spacetime which is not only
axisymmetric but also stationary.

Let $n^{\mu}$ be the unit timelike 4-vector orthogonal to 
the $t={\rm const.}$ hypersurface $\Sigma_t$
and oriented in the direction of increasing $t$,
\beqa
n_{\mu}=-N t_{,\mu}.
\label{eq:defn}
\eeqa
The lapse function $N$ is determined by the requirement
\beqa
n_{\mu} n^{\mu}=-1.
\label{eq:nn1}
\eeqa
The 3-metric induced by $g_{\mu \nu}$ on $\Sigma_t$ is given by
\beqa
h_{\mu \nu}=g_{\mu \nu}+n_{\mu} n_{\nu}.
\label{eq:defh}
\eeqa

Similarly, let $m_{\mu}$ be the unit spacelike 4-vector orthogonal to the 
$t={\rm const.}$ and $\varphi={\rm const.}$ 2-surface $\Sigma_{t\varphi}$
and oriented in the direction of increasing $\varphi$,
\beqa
m_{\mu}=M {h_{\mu}}^{\nu} \varphi_{,\nu}=M \varphi_{|\mu},
\label{eq:defm}
\eeqa
where the vertical stroke $|$ denotes the covariant derivative
associated with the 3-metric $h_{\mu \nu}$.
The coefficient $M$ is determined by 
\beqa
m_{\mu} m^{\mu}=1.
\label{eq:mm1}
\eeqa
The induced 2-metric on $\Sigma_{t\varphi}$ is given by 
\beqa
H_{\mu \nu}=h_{\mu \nu}-m_{\mu} m_{\nu}
=g_{\mu \nu}+n_{\mu} n_{\nu}-m_{\mu} m_{\nu}.
\label{eq:defH}
\eeqa
The covariant derivative associated with the 2-metric $H_{\mu \nu}$
is denoted by a double vertical stroke $\parallel$.
There is a relation between the determinants as
\beqa
\sqrt{-g}=N\sqrt{h}=N M \sqrt{H}.
\label{eq:detg}
\eeqa

Any 4-vector can be decomposed into its projection onto $\Sigma_{t\varphi}$,
the component parallel to $n_{\mu}$ and that to $m_{\mu}$.
The Killing vectors are decomposed as
\beqa
\eta^{\mu}&=&N n^{\mu}-N^{\mu}
=N n^{\mu} - M N^{\varphi} m^{\mu} - {N_{\Sigma}}^{\mu},
\label{eq:eta}
\\
\xi^{\mu}&=&M m^{\mu} - {M_{\Sigma}}^{\mu},
\label{eq:xi}
\eeqa
where the shift vector $N^{\mu}$ is (minus) the projection
of $\eta^{\mu}$ onto $\Sigma_{t}$, 
$M_{\Sigma}{}^{\mu}$ is (minus) the projection
of $\xi^{\mu}$ onto $\Sigma_{t\varphi}$, 
and $N_{\Sigma}{}^{\mu}$ is the projection
of $N^{\mu}$ onto $\Sigma_{t\varphi}$.
For our choice of coordinates, i.e., for $x^0=t$ and $x^3=\varphi$,
the component expressions for $n^{\mu}$, $n_{\mu}$, $m^{\mu}$ and $m_{\mu}$
are 
\beqa
n_{\mu}&=&(-N,0,0,0),
\label{eq:ncom1}
\\
n^{\mu}&=&\left(\frac{1}{N},\frac{N^{1}}{N},
\frac{N^{2}}{N},\frac{N^{\varphi}}{N}\right),
\label{eq:ncom2}
\\
m_{\mu}&=&(-M N^{\varphi},0,0,M),
\label{eq:mcom1}
\\
m^{\mu}&=&\left(0,\frac{{M_{\Sigma}}^{1}}{M},
\frac{{M_{\Sigma}}^{2}}{M},\frac{1}{M}\right).
\label{eq:mcom2}
\eeqa
Note that
$N_{\Sigma}{}^{\mu}=\left(0,N_{\Sigma}{}^{1},N_{\Sigma}{}^{2},0\right)$
and $N^{A}=N_{\Sigma}{}^{A}+N^{\varphi} M_{\Sigma}{}^{A}$.

The explicit component expressions of $g^{\mu \nu}$, $g_{\mu \nu}$, 
$h^{\mu \nu}$, and $h_{\mu \nu}$ are given by
\beqa
\left(
\begin{array}{cc}
g_{00} & g_{0j}\\
g_{i0} & g_{ij}
\end{array}
\right)
&=&
\left(
\begin{array}{cc}
N_{k} N^{k}-N^2~~
& -N_{j} \\
\\
-N_{i} & h_{ij}
\end{array}
\right),
\label{eq:g_}
\\
\nonumber\\
\left(
\begin{array}{cc}
g^{00} & g^{0j} \\
g^{i0} & g^{ij}
\end{array}
\right)
&=&
\left(
\begin{array}{cc}
\displaystyle{-\frac{1}{N^2}} & ~~\displaystyle{-\frac{N^{j}}{N^2}} \\
\\
\displaystyle{-\frac{N^{i}}{N^2}} & 
~~\displaystyle{h^{ij}-\frac{N^{i} N^{j}}{N^2}}
\end{array}
\right),
\label{eq:g^}
\\
\nonumber\\
\left(
\begin{array}{cc}
h_{AB} & h_{A3} \\
h_{3B} & h_{33}
\end{array}
\right)
&=&
\left(
\begin{array}{cc}
H_{AB} & -{M_{\Sigma}}_{A} \\
\\
-{M_{\Sigma}}_{B} &
~~ M^2+{M_{\Sigma}}_{A} {M_{\Sigma}}^{A}
\end{array}
\right),
\label{eq:h_}
\\
\nonumber\\
\left(
\begin{array}{cc}
h^{AB} & h^{A3} \\
h^{3B} & h^{33}
\end{array}
\right)
&=&
\left(
\begin{array}{cc}
\displaystyle{H^{AB}+\frac{{M_{\Sigma}}^{A} {M_{\Sigma}}^{B}}{M^2}} & 
\displaystyle{\frac{{M_{\Sigma}}^{A}}{M^2}} \\
\\
\displaystyle{\frac{{M_{\Sigma}}^{B}}{M^2}} & 
\displaystyle{\frac{1}{M^2}}
\end{array}
\right).
\label{eq:h^}
\eeqa
where $i, j, k=1,2,3$ and $A, B=1,2$.
We can express the 4-metric $g_{\mu\nu}$
in terms of $N$, $N^{\varphi}$, $N_{\Sigma}{}^{A}$,
$M$, $M_{\Sigma}{}^{A}$ and $H_{AB}$ as
\beqa
g_{\mu \nu} dx^{\mu} dx^{\nu}
&=&-\left[N^2-M^2 \left(N^{\varphi}\right)^2-
N_{\Sigma}{}_{A} N_{\Sigma}{}^{A}\right] dt^2
-2 \left(M^2 N^{\varphi}-N_{\Sigma}{}^{A} M_{\Sigma}{}_{A}\right) dt d\varphi
\nonumber\\
&-&2 N_{\Sigma}{}_{A} dt dx^A
-2 M_{\Sigma}{}_{A} d\varphi dx^A
+H_{AB} dx^{A} dx^{B}
+\left(M^2+M_{\Sigma}{}_{A} M_{\Sigma}{}^{A}\right) d\varphi^2,
\label{eq:metric}
\eeqa
where the functions $N$, $N^{\varphi}$, $N_{\Sigma}{}^{A}$,
$M$, $M_{\Sigma}{}^{A}$ and $H_{AB}$ depend only
on the coordinate $(x^{1}, x^{2})$.
Since we only assume that physical quantities
are independent of $x^{0}=t$ and $x^{3}=\varphi$,
the metric $g_{\mu \nu}$ in Eq.~(\ref{eq:metric})
has some freedom in the choice of coordinates.
We will leave the coordinate freedom unspecified.
In Sec.~\ref{sec:GScov} the covariant GS equation will be given
as an equation projected onto $\Sigma_{t\varphi}$.

\section{Physical quantities from flux function $\Psi$}
\label{sec:inv}

Provided that the metric $g_{\mu\nu}$ is given and
the conserved quantities $E(\Psi)$, $L(\Psi)$, $\Omega(\Psi)$, 
$C(\Psi)$ and $S(\Psi)$ are given as functions of $\Psi$,
all the physical quantities can be evaluated once the
(effectively 2-dimensional) configuration of the flux function
$\Psi$ is known (see \cite{c86b} for the circular case).
In this section, we explicitly demonstrate this fact
for the most general case of noncircular spacetimes.

\subsection{Fluid 4-velocity $u^{\mu}$}
First let us consider the fluid 4-velocity $u^{\mu}$.
It is useful to prepare two vectors $\eta^{\mu} + \Omega \xi^{\mu}$
and $\xi^{\mu} + \Theta \eta^{\mu}$ constructed from two Killing vectors
$\eta^{\mu}$ and $\xi^{\mu}$,
and to make them orthogonal to each other
$(\eta^{\mu}+\Omega \xi^{\mu})(\xi_{\mu}+\Theta \eta_{\mu})=0$
by taking
\beqa
\Theta=-\frac{\xi_{\nu}(\eta^{\nu}+\Omega \xi^{\nu})}
{\eta_{\mu}(\eta^{\mu}+\Omega \xi^{\mu})}
=-\frac{g_{03}+\Omega g_{33}}{g_{00}+\Omega g_{03}}.
\label{eq:Theta}
\eeqa
Then we can decompose the fluid 4-velocity in the coordinate bases as
\beqa
u^{\mu}=u_{\eta} (\eta^{\mu} + \Omega \xi^{\mu})
+u_{\xi} (\xi^{\mu} + \Theta \eta^{\mu})+\tilde u_{\Sigma}{}^{\mu},
\label{eq:uinv}
\eeqa
where $\eta^{\mu} + \Omega \xi^{\mu}=(1,0,0,\Omega)$,
$\xi^{\mu} + \Theta \eta^{\mu}=(\Theta,0,0,1)$
and $\tilde u_{\Sigma}{}^{\mu}=(0,u^{1},u^{2},0)$
in the component expressions,
and hence
$\tilde u_{\Sigma}{}^{\mu} n_{\mu}=\tilde u_{\Sigma}{}^{\mu} m_{\mu}=0$
from Eqs.~(\ref{eq:ncom1}) and (\ref{eq:mcom1}), and
\beqa
u^{0}&=&u_{\eta}+\Theta u_{\xi},
\label{eq:u0}
\\
u^{3}&=&u_{\xi}+\Omega u_{\eta}.
\label{eq:u3}
\eeqa
The decomposition in Eq.~(\ref{eq:uinv})
is not conforming to the spirit of the $(2+1)+1$
formalism but makes it easy to obtain the coefficients $u_{\eta}$
and $u_{\xi}$ as shown below.

{}From Eqs.~(\ref{eq:F31}) and (\ref{eq:F23}),
the term $\tilde u_{\Sigma}{}^{\mu}$ is given by
\beqa
\tilde u_{\Sigma}{}^{\mu}
=\frac{1}{N M C \rho} \epsilon^{\mu \nu} \Psi_{,\nu},
\label{eq:tuperp}
\eeqa
where the antisymmetric tensor $\epsilon^{\mu \nu}$ is defined by
\beqa
\epsilon^{\mu \nu}=\epsilon^{\mu \nu \alpha \beta} n_{\alpha} m_{\beta}.
\label{eq:epsi2}
\eeqa
With Eqs.~(\ref{eq:Bu}) -- (\ref{eq:defL}), (\ref{eq:eta}) and (\ref{eq:xi}), 
the coefficients $u_\eta$ and $u_\xi$ are expressed as
\beqa
u_{\eta}&=&\frac{E-\Omega L}{G_{\eta} \mu}-\frac{\tilde N_{\Sigma}}{G_{\eta}}
=\frac{D}{G_{\eta} \mu}-\frac{\tilde N_{\Sigma}}{G_{\eta}},
\label{eq:uomega}
\\
u_{\xi}&=&-\frac{(L-\Theta E)}
{G_{\xi} \mu}
\left(\frac{4\pi \mu}{G_{\eta} C^2 \rho}\right)
\left(1-\frac{4\pi \mu}{G_{\eta} C^2 \rho}\right)^{-1}
+\frac{\tilde M_{\Sigma}}{G_{\xi}},
\label{eq:utheta}
\eeqa
where $G_\eta$ and $G_\xi$ are defined by
\beqa
G_{\eta}&=&-(\eta_{\mu} + \Omega \xi_{\mu})(\eta^{\mu} + \Omega \xi^{\mu})
=-(g_{00}+2\Omega g_{03}+\Omega^{2} g_{33}),
\label{eq:Geta}
\\
G_{\xi}&=&(\xi_{\mu} + \Theta \eta_{\mu})(\xi^{\mu} + \Theta \eta^{\mu})
=g_{33}+2\Theta g_{03}+\Theta^{2} g_{00},
\label{eq:Gxi}
\eeqa
and $\tilde N_\Sigma$ and $\tilde M_\Sigma$ are defined by
\beqa
\tilde N_{\Sigma}&=&\tilde u_{\Sigma}{}^{\mu}
\left(N_{\Sigma}{}_{\mu}+\Omega M_{\Sigma}{}_{\mu}\right),
\label{eq:tN}
\\
\tilde M_{\Sigma}&=&\tilde u_{\Sigma}{}^{\mu}
\left(M_{\Sigma}{}_{\mu}+\Theta N_{\Sigma}{}_{\mu}\right).
\label{eq:tM}
\eeqa
Note that, using Eq.~(\ref{eq:tuperp}),
 $\tilde N_\Sigma$ and $\tilde M_\Sigma$ are
expressed as
\begin{eqnarray}
\tilde N_\Sigma=\frac{1}{NMC\rho}
\epsilon^{AB}\Psi_{\parallel B}(N_{\Sigma}{}_A+\Omega M_{\Sigma}{}_A)\,,
\label{eq:tNexp}
\\
\tilde M_\Sigma=\frac{1}{NMC\rho}
\epsilon^{AB}\Psi_{\parallel B}(M_{\Sigma}{}_A+\Theta N_{\Sigma}{}_A)\,.
\label{eq:tMexp}
\end{eqnarray}
Note also that 
\begin{eqnarray}
{M_{\rm Alf}}^2&:=&\frac{4\pi \mu}{G_{\eta} C^2 \rho}
\nonumber\\
&=&\frac{4 \pi \mu \rho}{B^2} \left(G_{\xi} u_{\xi}{}^{2}+
\tilde u_{\Sigma}{}^{A} \tilde u_{\Sigma}{}_{A}
-2 u_{\xi} \tilde M_{\Sigma}+\frac{\tilde N_{\Sigma}{}^2}{G_{\eta}}\right),
\label{eq:Malf}
\end{eqnarray}
is the square of the effective Alfv$\acute{\rm e}$n Mach number $M_{\rm Alf}$
(where the second equality follows from Eq.~(\ref{eq:BB}) below).
At the Alfv$\acute{\rm e}$n point $M_{\rm Alf}=1$,
the numerator $L-\Theta E$ should vanish to keep the velocity
$u_{\xi}$ finite \cite{c86a}.

Thus, from Eqs.~(\ref{eq:uinv}), (\ref{eq:uomega}), (\ref{eq:utheta}) 
and (\ref{eq:tuperp}), given the metric $g_{\mu\nu}$ and
the conserved functions $E(\Psi)$, $L(\Psi)$, $\Omega(\Psi)$, 
$C(\Psi)$ and $S(\Psi)$, 
the fluid 4-velocity $u^{\mu}$ can be obtained from the 
flux function $\Psi$ and its first derivatives $\Psi_{,A}$
if the density $\rho$ and the enthalpy $\mu$
are additionally known. The expression for $\rho$ 
will be given in the next subsection. The enthalpy $\mu$ 
is then determined as a function of $\rho$ and $\Psi$
as will be discussed also in the next subsection.

Once the components of $u^\mu$ are known,
the $(2+1)+1$ decomposition of the fluid 4-velocity 
is easily performed. With the help of Eqs.~(\ref{eq:uinv}), (\ref{eq:eta}) 
and (\ref{eq:xi}), we have
\beqa
u^{\mu}=u_{n} n^{\mu}
+u_{m} m^{\mu}
+ {u_{\Sigma}}^{\mu},
\label{eq:u2+1+1}
\eeqa
where
\beqa
u_{n}&=&N (u_{\eta} + \Theta u_{\xi}),
\label{eq:un}
\\
u_{m}&=&M \left[\left(\Omega-N^{\varphi}\right) u_{\eta}
+\left(1-N^{\varphi}\Theta\right) u_{\xi}\right],
\label{eq:um}
\\
{u_{\Sigma}}^{\mu}&=&\tilde u_{\Sigma}{}^{\mu}
-(u_{\eta}+\Theta u_{\xi}) {N_{\Sigma}}^{\mu}
-(u_{\xi} + \Omega u_{\eta}) {M_{\Sigma}}^{\mu}.
\label{eq:uperp}
\eeqa

\subsection{Density $\rho$ and other thermodynamical quantities
$p$, $\epsilon$, $\mu$ and $T$}
The pressure $p$, the internal energy $\epsilon$, 
the enthalpy $\mu$ and the temperature $T$
are functions of the density $\rho$ and the entropy $S$
from Eqs.~(\ref{eq:defmu}), (\ref{eq:1law}) and (\ref{eq:eos}).
Hence, given $S$ as a function of $\Psi$, the only remaining quantity
to be known is the density $\rho$.

The density $\rho$ is determined by the normalization of
the 4-velocity $u_{\mu} u^{\mu} = -G_{\eta} {u_{\eta}}^2
+G_{\xi} {u_{\xi}}^2+
\tilde u_{\Sigma}{}_{A}\tilde u_{\Sigma}{}^{A}
+2 u_{\eta} \tilde u_{\Sigma}{}^{\mu}\left(\eta_{\mu}+\Omega \xi_{\mu}\right)
+2 u_{\xi} \tilde u_{\Sigma}{}^{\mu}\left(\xi_{\mu}+\Theta \eta_{\mu}\right)
=-1$, that is,
\beqa
-\frac{(E-\Omega L)^2}{G_{\eta} \mu^2}
+\frac{(4\pi)^2 (L-\Theta E)^2}
{{G_{\xi} G_{\eta}}^2 C^4 \rho^2}
\left(1-\frac{4\pi \mu}{G_{\eta} C^2 \rho}\right)^{-2}
+\frac{H^{A B} \Psi_{,A} \Psi_{,B}}{N^2 M^2 C^2 \rho^2}
+\frac{\tilde N_{\Sigma}{}^{2}}{G_{\eta}}-
\frac{\tilde M_{\Sigma}{}^{2}}{G_{\xi}}
=-1.
\label{eq:uu1}
\eeqa
This equation is what is called the wind equation (see \cite{c86b}
for the circular case). Note that $\rho$ contains the first-order
derivatives $\Psi_{,A}$ through this equation.

\subsection{Magnetic field $B^{\mu}$}
The magnetic field is also calculated from the flux function.
With Eqs.~(\ref{eq:Bu}), (\ref{eq:u2+1+1}), (\ref{eq:eta}) and (\ref{eq:xi}), 
the $(2+1)+1$ decomposition of the magnetic field is given by
\beqa
B^{\mu}=B_n n^{\mu}+B_{m} m^{\mu}+B_{\Sigma}{}^{\mu},
\label{eq:B2+1+1}
\eeqa
where
\beqa
B_{n}&=&C\rho N\left[\left(G_{\eta} u_{\eta}+\tilde N_{\Sigma}\right)
\left(u_{\eta}+\Theta u_{\xi}\right)-1\right],
\label{eq:Bn}
\\
B_{m}&=&C\rho M\left[\left(G_{\eta} u_{\eta}+\tilde N_{\Sigma}\right)
\left\{
\left(\Omega-N^{\varphi}\right) u_{\eta}
+\left(1-N^{\varphi}\Theta\right) u_{\xi}
\right\}
+N^{\varphi}-\Omega\right],
\label{eq:Bm}
\\
B_{\Sigma}{}^{\mu}&=&C\rho\left[
\left(G_{\eta} u_{\eta}+\tilde N_{\Sigma}\right) u_{\Sigma}{}^{\mu}
+N_{\Sigma}{}^{\mu}+\Omega M_{\Sigma}{}^{\mu}\right].
\label{eq:Bperp}
\eeqa
The magnetic strength is given by
\beqa
B^{2}=B^{\mu} B_{\mu}=
C^2 \rho^2 \left[\left(G_{\eta} {u_{\eta}}+\tilde N_{\Sigma}\right)^2
-G_{\eta} \right]
=C^2 \rho^2 \left[G_{\eta}\left(G_{\xi} {u_{\xi}}^2
+\tilde u_{\Sigma}{}_{A} \tilde u_{\Sigma}{}^{A}
-2 u_{\xi} \tilde M_{\Sigma}\right)+\tilde N_{\Sigma}{}^2\right].
\label{eq:BB}
\eeqa
The $(2+1)+1$ decomposition of the energy-momentum tensor
is given in Appendix~\ref{sec:emt2+1+1}.

\subsection{Electric current $J^{\mu}$}\label{subsec:J}
Let us consider the following components of the electric current,
\beqa
J^{0}=\frac{1}{4\pi N M \sqrt{H}}\left(N M \sqrt{H} F^{0A}\right)_{,A}
=\frac{1}{4\pi N M}\left(N M F^{0A}\right)_{||A},
\label{eq:J0}
\\
J^{3}=\frac{1}{4\pi N M \sqrt{H}}\left(N M \sqrt{H} F^{3A}\right)_{,A}
=\frac{1}{4\pi N M}\left(N M F^{3A}\right)_{||A}.
\label{eq:J3}
\eeqa
The field strength tensor components
appearing in the above equations 
are also expressed in terms of the flux function $\Psi$ as
\beqa
F^{0A}&=&\left(g^{00} g^{AB} - g^{0B} g^{A0}\right) \Omega \Psi_{,B}
+\left(g^{0B} g^{A3} - g^{03} g^{AB}\right) \Psi_{,B}
+\left(g^{01} g^{A2} - g^{02} g^{A1}\right) F_{12}
\nonumber\\
&=&-\frac{1}{N^2} \left[H^{AB}+
\frac{M_{\Sigma}^A M_{\Sigma}^B}{M^2}\right] \Omega \Psi_{,B}
+\frac{1}{N^2} \left[N^{\varphi} H^{AB} - 
\frac{M_{\Sigma}^A {N_{\Sigma}}^{B}}{M^2}\right] \Psi_{,B}
\nonumber\\
&&-\left[\left(\frac{{N_{\Sigma}}^{B} + N^{\varphi} M_{\Sigma}^{B}}{N^2}\right)
{\epsilon_B}^A
+\frac{M_{\Sigma}^A {N_{\Sigma}}^B M_{\Sigma}^C \epsilon_{BC}}{N^2 M^2}
\right]
\frac{F_{12}}{\sqrt{H}},
\label{eq:Fu0A}
\\
F^{3A}&=&\left(g^{30} g^{AB} - g^{3B} g^{A0}\right) \Omega \Psi_{,B}
+\left(g^{3B} g^{A3} - g^{33} g^{AB}\right) \Psi_{,B}
+\left(g^{31} g^{A2} - g^{32} g^{A1}\right) F_{12}
\nonumber\\
&=&-\frac{1}{N^2} \left[N^{\varphi} H^{AB}
-\frac{{N_{\Sigma}}^{A} M_{\Sigma}^{B}}{M^2}\right] \Omega \Psi_{,B}
-\left[\left(\frac{1}{M^2}-\left(\frac{N^{\varphi}}{N}\right)^2\right)
H^{AB} - \frac{{N_{\Sigma}}^{A} {N_{\Sigma}}^{B}}{N^2 M^2}\right]
\Psi_{,B}
\nonumber\\
&&+\left[
\left\{
\left(\frac{1}{M^2}-\left(\frac{N^{\varphi}}{N}\right)^2\right) M_{\Sigma}^{B} 
-\frac{N^{\varphi} {N_{\Sigma}}^{B}}{N^2} 
\right\}{\epsilon_{B}}^{A}
+\frac{{N_{\Sigma}}^{A} {N_{\Sigma}}^{B} M_{\Sigma}^{C} \epsilon_{BC}}{N^2 M^2}
\right]
\frac{F_{12}}{\sqrt{H}},
\label{eq:Fu3A}
\eeqa
where $F_{12}$ is expressed as
\beqa
\frac{F_{12}}{\sqrt{H}}=C N M \rho (u^3 - \Omega u^0)
=C N M \rho u_{\xi}
\left(1 - \Omega \Theta \right).
\label{eq:F12H}
\eeqa
In the above, the first equality follows from Eq.~(\ref{eq:F12}),
and the second from Eqs.~(\ref{eq:u0}) and (\ref{eq:u3}).
Thus, $J^0$ and $J^3$ are expressed in terms of $\Psi$ and its
first and second derivatives.

\subsection{Auxiliary quantity $\Lambda$}
We also need to evaluate the auxiliary quantity $\Lambda$
defined by Eq.~(\ref{eq:Lambdadef}), that is,
\beqa
\Lambda=\frac{1}{4\pi} (u_0 B_3 - u_3 B_0)\,.
\eeqa
{}From the expression of $B^\mu$ given by Eq.~(\ref{eq:Bu}),
we have
\beqa
\Lambda=-\frac{1}{4\pi}
C\rho\left[u_{0}\left(g_{03}+\Omega g_{33}\right)
-u_{3}\left(g_{00}+\Omega g_{03}\right)\right]\,.
\eeqa
Using the component expressions of $u^\mu$ given by Eq.~(\ref{eq:uinv}),
this is rewritten as
\beqa
\Lambda&=&-\frac{1}{4\pi}
C\rho\left[u_{\xi} \left(g_{03}{}^{2}-g_{00} g_{33}\right)
\left(1-\Omega\Theta\right)
+\tilde u_{\Sigma}{}^{\mu} \eta_{\mu} \left(g_{03}+\Omega g_{33}\right)
-\tilde u_{\Sigma}{}^{\mu} \xi_{\mu} \left(g_{00}+\Omega g_{03}\right)
\right]
\nonumber\\
&=&-\frac{1}{4\pi}
C\rho\left[u_{\xi} \left(g_{03}{}^{2}-g_{00} g_{33}\right)
\left(1-\Omega\Theta\right)
+\tilde M_{\Sigma}\left(g_{00}+\Omega g_{03}\right)\right],
\eeqa
where the second line follows from 
Eqs.~(\ref{eq:eta}), (\ref{eq:xi}), (\ref{eq:Theta})
and (\ref{eq:tM}).
The above form is sufficient for $\Lambda$ to be obtained from the 
flux function, but it can be further simplified if we
use Eqs.~(\ref{eq:Theta}) and (\ref{eq:Gxi}). {}From
these equations, we find
\[
(1-\Omega \Theta)\left(g_{00} g_{33}-{g_{03}}^2\right)=G_{\xi}
\left(g_{00}+\Omega g_{03}\right)\,.
\]
Therefore, we obtain
\beqa
\Lambda=\frac{1}{4\pi}
(u_0 B_3 - u_3 B_0)=
\frac{1}{4\pi} C \rho \left(G_{\xi} u_{\xi}-\tilde M_{\Sigma}\right)
\left(g_{00}+\Omega g_{03}\right)\,.
\label{eq:Lambda}
\eeqa

\section{Grad-Shafranov equation in the covariant form}
\label{sec:GScov}

Now we are ready to show that the GS equation~(\ref{eq:GScom})
is indeed a second-order differential equation for the flux function
$\Psi$. At the same time, following the spirit of the
$(2+1)+1$ formalism, we express the GS equation in the covariant 
form with respect to the geometry of $\Sigma_{t\varphi}$.

The covariant expression for the GS equation is readily obtained as
\beqa
J^{3}-\Omega J^{0}
+\frac{1}{N M C} \epsilon^{AB} (\mu {u_{\Sigma}}_{A})_{||B}
-\rho (u_{\eta} + \Theta u_{\xi}) \left[E'-\Lambda (C \Omega)'\right]
+\rho (u_{\xi} + \Omega u_{\eta}) \left[L'-\Lambda C'\right]
+\rho T S'=0,
\label{eq:GS}
\eeqa
where we have replaced $\sqrt{-g}$, $u^0$ and $u^3$ in the 
original GS equation~(\ref{eq:GScom}) by their $(2+1)+1$ type
expressions~(\ref{eq:detg}), (\ref{eq:u0}) and (\ref{eq:u3}),
respectively,
and, as before, a double vertical stroke $\parallel$ denotes
the covariant differentiation
with respect to the 2-metric $H_{AB}$.

In the previous section, we have seen that
 $u_{\eta}$, $u_{\xi}$, $u_{\Sigma}{}^{A}$ and $\Theta$
(section \ref{sec:inv} A), 
$\rho$, $\mu$ and $T$ (section \ref{sec:inv} B), 
$J^0$ and $J^3$ (section \ref{sec:inv} D),
and $\Lambda$ (section \ref{sec:inv} E) are all expressed 
in terms of $\Psi$ and its derivatives, given the conserved functions
 $E(\Psi)$, $L(\Psi)$, $\Omega(\Psi)$, $C(\Psi)$ and $S(\Psi)$,
and the metric $g_{\mu\nu}$. 
In particular, we have seen that $J^0$ and $J^3$ contain 
the second-order derivatives of $\Psi$, while $\rho$ (hence $\mu$) 
as well as $u_{\Sigma}{}_A$ contain the first-order derivatives of $\Psi$.
Thus, the GS equation~(\ref{eq:GS}) is a second-order, non-linear
differential equation for $\Psi$, where the first three terms
contain the second-order derivatives.

\subsection{No toroidal field limit}
\label{sec:noT}
{}From Eqs.~(\ref{eq:F31}) -- (\ref{eq:F12}),
we find that the toroidal field and the meridional flow vanish if
$|C|\to \infty$.
Here note that $u^3-\Omega u^0=u_{\xi}(1-\Omega \Theta)
\propto C^{-2} \to 0$
in Eq.~(\ref{eq:F12}) from Eqs.~(\ref{eq:u0}), (\ref{eq:u3}) and 
(\ref{eq:utheta})
(and hence $\Omega$ coincides with the angular velocity
$d\varphi/dt=u^3/u^0=\Omega$).
In the absence of the toroidal field and the meridional flow,
a spacetime is circular.
The circular limit is expressed as \cite{gb93}
\beqa
N_{\Sigma}{}^{A}\to 0,\quad
M_{\Sigma}{}^{A}\to 0.
\label{eq:circ}
\eeqa
Therefore, in the $|C|\to \infty$ limit,
the GS equation~(\ref{eq:GS}) reduces to
\beqa
J^{3}-\Omega J^{0}-\rho u_{\eta}
\left[E'-\Omega L'-C \Lambda \Omega'\right]+\rho T S'=0,
\label{eq:GSc}
\eeqa
where the density is determined by
${\left(E-\Omega L\right)^2}/{G_{\eta} \mu^2}=
G_{\eta} {u_{\eta}}^{2}=1$ from Eq.~(\ref{eq:uu1}),
and
\beqa
C \Lambda=-\frac{L-\Theta E}{G_{\eta}}
\left(g_{00}+\Omega g_{03}\right)
\left(1-\frac{4\pi\mu}{G_{\eta} C^2 \rho}\right)^{-1}.
\label{eq:clambda}
\eeqa
Here we regard $|C|\to \infty$ as the limit of a sequence of
models with $|C|<\infty$.
The last term $\left[1-(4\pi\mu)/(G_{\eta} C^2 \rho)\right]^{-1}$
in Eq.~(\ref{eq:clambda}) can be neglected if the density $\rho$ is finite.
However, in the case when there is a surface with $\rho=0$ like a star
and the flux function $\Psi$ is not constant on that surface,
the last term $\left[1-(4\pi\mu)/(G_{\eta} C^2 \rho)\right]^{-1}$
diverges near the surface. Unless one can fine-tune the
term $L-\Theta E$ so that its zero point cancels this divergence,
which seems unlikely to be possible, we should demand
 the rigid rotation $\Omega'=0$ in Eq.~(\ref{eq:GSc}).
This is consistent with Bonazzola et al. \cite{b93,b95}.
Note that if the flux function $\Psi$ is constant on the 
$\rho=0$ surface,
we may find $C$ that satisfies $C^2 \rho \to \infty$ on the surface.



\subsection{No poloidal field limit}
The poloidal field vanishes if we let $\Psi \to \delta \hat \Psi$
and take a limit $\delta\to 0$, as we can see from Eqs.~(\ref{eq:F31})
and (\ref{eq:F23}).
In this process we relabel the flow lines by $\hat \Psi$
and replace the conserved quantities as
$E\to E(\hat \Psi)$, $L\to L(\hat \Psi)$,
$\Omega\to \Omega(\hat \Psi)$,
$C\to C(\hat \Psi)$
and $S\to S(\hat \Psi)$.
In the limit $\delta \to 0$, the meridional flow vanishes 
$\tilde u_{\Sigma}{}^{A}\to 0$ from 
Eq.~(\ref{eq:tuperp}).
Then we can show that the spacetime is circular 
as expressed in Eq.~(\ref{eq:circ}) \cite{o02}.
Therefore, in the $\delta \to 0$ limit, the GS equation~(\ref{eq:GS})
reduces to
\beqa
\left(u_{\eta}+\Theta u_{\xi}\right)
\left[E'-\Lambda \left(C \Omega\right)'\right]
-\left(u_{\xi}+\Omega u_{\eta}\right)
\left[L'-\Lambda C'\right]
-T S'=0,
\eeqa
where primes now denote differentiation with respect to $\hat \Psi$.
This is an algebraic equation.
Here we regard $\Psi\to 0$ as the limit of a sequence
of models with $\Psi\ne 0$.
If $\Psi$ is exactly zero,
the transfield components
of the Euler equations (\ref{eq:euler}) are satisfied
regardless of the GS equation (see Sec.~\ref{sec:gscom}).
Therefore there may exist 'isolated' solutions
which cannot be obtained by the limit discussed here.

\subsection{No magnetic field limit}
There are two limits for configurations with no magnetic field.
The first way to obtain such configurations is to let 
$\Psi \to \delta_1 \hat \Psi$ and $C \to \hat C/\delta_2$
and take the limit $\delta_1 \to 0$ and $\delta_2 \to 0$,
as we can see from Eqs.~(\ref{eq:F31}) -- (\ref{eq:F12}).
Here note that $u^3-\Omega u^0=u_{\xi}(1-\Omega \Theta)
\propto C^{-2} \to 0$ in Eq.~(\ref{eq:F12})
from Eqs.~(\ref{eq:u0}), (\ref{eq:u3}) and (\ref{eq:utheta})
(and hence $\Omega$ coincides with the angular velocity
$d\varphi/dt=u^3/u^0=\Omega$).
In this process we relabel the flow lines by $\hat \Psi$
and replace the conserved quantities as
$E\to E(\hat \Psi)$, $L\to L(\hat \Psi)$,
$\Omega\to \Omega(\hat \Psi)$,
$C\to C(\hat \Psi)$
and $S\to S(\hat \Psi)$.
In the limit $\delta_1 \to 0$ and $\delta_2 \to 0$, 
the meridional flow vanishes $\tilde u_{\Sigma}{}^{A}\to 0$ from 
Eqs.~(\ref{eq:F31}) and (\ref{eq:F23}),
so that the spacetime becomes circular as in Eq.~(\ref{eq:circ}).
Therefore, the GS equation~(\ref{eq:GS}) reduces to 
\beqa
u_{\eta}\left[\left(E-\Omega L\right)'-
\Omega'\left(C \Lambda-L\right)\right]
-T S'=0,
\label{eq:GSno1}
\eeqa
where primes denote differentiation with respect to $\hat \Psi$,
$C \Lambda$ is given by Eq.~(\ref{eq:clambda}),
and the density is determined from Eq.~(\ref{eq:uu1}) as 
\beqa
\frac{(E-\Omega L)^2}{G_{\eta} \mu^2}
=G_{\eta} u_{\eta}{}^2=1\,.
\label{eq:nouu1}
\eeqa

Let us see the relation between
this limit ($\Psi \to 0$ and $|C| \to \infty$)
and the case of a rotating star \cite{b70,b93}.
{}From Eq.~(\ref{eq:nouu1})
we have the Bernouilli's equation for a rotating fluid as
\beqa
\ln \mu-\ln u_{\eta}-\ln \left(E-\Omega L\right)=0.
\eeqa
{}From the definition of $L$, Eq.~(\ref{eq:defL2}), 
we have $C \Lambda-L=-\mu u_3=-u_{3} u_{\eta} (E-\Omega L)$.
Then for an isentropic star $S'=0$,
the GS equation in Eq.~(\ref{eq:GSno1}) is written as
\beqa
u_{3} u_{\eta}=\frac{d}{d\Omega}
\left[-\ln (E-\Omega L)\right]\,,
\label{eq:rotation}
\eeqa
where we have used the fact that
$-\ln (E-\Omega L)$ can be regarded as a function of $\Omega$,
since $E$, $L$ and $\Omega$ are functions of $\hat \Psi$ only.
Equation~(\ref{eq:rotation}) is the well-known integrability condition 
for a rotating fluid \cite{b70,b93}.

Note that
if we regard $|C|\to \infty$ as the limit of a sequence
of models with $|C|<\infty$, we should demand the rigid rotation 
$\Omega'=0$,
as discussed in Sec.~\ref{sec:noT}.

The second way to obtain configurations with no magnetic field
is to let $\Psi \to \delta \hat \Psi$
and $C \to \delta \hat C$
and take the limit $\delta\to 0$.
In this process we relabel the flow lines by $\hat \Psi$
and replace the conserved quantities as
$E\to E(\hat \Psi)$, $L\to L(\hat \Psi)$,
$\Omega\to \Omega(\hat \Psi)$,
$\hat C\to \hat C(\hat \Psi)$
and $S\to S(\hat \Psi)$.
As we can see from Eqs.~(\ref{eq:F31}) and (\ref{eq:F23}),
there exists a meridional flow in this case.
The toroidal field vanishes since in Eq.~(\ref{eq:F12})
$u^{3}-\Omega u^{0}=u_{\xi} (1-\Omega \Theta)$
and $u_{\xi} \to (L-\Theta E)/G_{\xi} \mu+\tilde M_{\Sigma}/G_{\xi}$
in the limit $\delta \to 0$, from Eqs.~(\ref{eq:u0}), (\ref{eq:u3}) 
and (\ref{eq:utheta}).
The GS equation in Eq.~(\ref{eq:GS}) reduces to
\beqa
-\rho u_{\eta} \left(E'-\Omega L'\right)
+\rho u_{\xi} \left(L'-\Theta E'\right)
+\frac{1}{NM \hat C}
\epsilon^{AB} \left(\mu u_{\Sigma}{}_{A}\right)_{||B}
+\rho T S'=0,
\eeqa
where primes denote differentiation with respect to $\hat \Psi$.
One can introduce a yet new flux function $\tilde\Psi$ by
$d\tilde\Psi =d\hat\Psi/\hat C$ to absorb the function $\hat C$
into the definition of the new flux function $\tilde\Psi$.
The resulting equation may be directly obtained from the Euler equations
for a perfect fluid.

\subsection{Newtonian limit}
In the Newtonian limit, all physical quantities are expanded 
in power series of the typical fluid velocity \cite{w72}.
The metric reduces to
\beqa
g_{\mu \nu} dx^{\mu} dx^{\nu}
=-(1+2\phi)dt^2+(1-2\phi)\left(dZ^2+dR^2+R^2 d\varphi^2\right),
\label{eq:metricN}
\eeqa
where the Newtonian potential $\phi$ is of order $O(v^2)$,
and we adopt the cylindrical coordinate $(t,Z,R,\varphi)$.
We denote the 3-dimensional velocity by
\beqa
v^{i}:=\frac{u^{i}}{u^{0}}=\frac{dx^{i}}{dt},
\label{eq:3velo}
\eeqa
where 
\beqa
u^{0}=1-\phi+\frac{v^2}{2},
\eeqa
and $v^2=v^{i} v_{i}$.
We regard the internal energy $\epsilon$ and the pressure $p$ 
to be $O(v^2)$.
To make the energy density of the electromagnetic field $O(v^2)$,
we demand 
\beqa
B^{i}\sim O(v)\,,\quad
B^{0}\sim O(v^2)\,,\quad
\Psi \sim O(v)\,,\quad
\Omega \sim O(v)\,,
\eeqa
from Eqs.~(\ref{eq:FuB}) and (\ref{eq:F03}) -- (\ref{eq:F12}).

{}From Eqs.~(\ref{eq:F31}) -- (\ref{eq:F12}),
we find 
\beqa
B^{A}&=&C\rho v^{A},
\\
B^{\hat\varphi}&=&C\rho\left(v^{\hat\varphi}-R\Omega\right),
\eeqa
where $B^{\hat\varphi}:=R\,B^3$ and $v^{\hat\varphi}:=R\,v^3$.
{}From Eqs.~(\ref{eq:defL2}), (\ref{eq:Lambda}) and (\ref{eq:defD}),
we also have
\beqa
L&=&R\,v^{\hat\varphi}+C\Lambda
=R \left(v^{\hat\varphi}-\frac{CB^{\hat\varphi}}{4\pi}\right) \sim O(v),
\label{eq:LNew}
\\
D-1&=&\epsilon +\frac{p}{\rho}+\frac{v^2}{2}+\phi
-R \,\Omega\, v^{\hat\varphi} \sim O(v^2).
\eeqa
These results are to be compared with the Newtonian results
(note the correspondences between our notation
and that of \cite{l86} as
$\Omega \leftrightarrow G$,
$C \leftrightarrow 4\pi/F$,
$L \leftrightarrow -H/F$,
and $D-1 \leftrightarrow J$).

Let us obtain the Newtonian GS equation, which is of order $O(v)$.
First consider the $J^3$ and $-\Omega J^0$ terms in Eq.~(\ref{eq:GS}).
In the Newtonian order, these terms may be evaluated
on the flat background with $N=1$ and $M=R^2$.
Then $-\Omega J^0$ is found to be $O(v^3)$, and
the term $J^3$ is given by
\beqa
J^3=-\frac{1}{4\pi N M}\left(\frac{N}{M} H^{AB} \Psi_{,B}\right)_{||A}
=-\frac{1}{4\pi R^2} \Delta^{*} \Psi,
\label{eq:J3N}
\eeqa
where
\beqa
\Delta^{*}=R\frac{\partial}{\partial R}\frac{1}{R}\frac{\partial}{\partial R}
+\frac{\partial^2}{\partial Z^2}.
\label{eq:defd*}
\eeqa
Next consider the $\epsilon^{AB}(\mu u_\Sigma{}_A)_{\parallel B}$ term
in Eq.~(\ref{eq:GS}).
By using Eqs.~(\ref{eq:uperp}) and (\ref{eq:tuperp}), we have
\beqa
\frac{1}{N M C} \epsilon^{AB} \left(\mu u_{\Sigma}{}_{A}\right)_{||B}
=\frac{1}{NMC} \left(\frac{\mu}{NMC\rho} \Psi_{,A}\right)_{||B} H^{AB}
=\frac{1}{4\pi R^2}\left[\frac{4\pi}{C^2\rho}
\Delta^{*} \Psi
+\frac{4\pi}{C}\nabla \left(\frac{1}{C\rho}\right)\cdot
\nabla \Psi\right].
\label{eq:upN}
\eeqa
Finally consider the terms proportional to $\rho$ in Eq.~(\ref{eq:GS}).
To the lowest order,
from Eqs.~(\ref{eq:Theta}), (\ref{eq:uomega}), (\ref{eq:utheta})
and (\ref{eq:Lambda}), we have 
\beqa
\Theta&=&R^2 \Omega \sim O(v),
\\
u_{\eta}&=&1\sim O(1), 
\\
u_{\xi}&=&-\frac{(L-R^2 \Omega)}{R^2}
\left(\frac{4\pi}{C^2\rho}\right)
\left(1-\frac{4\pi}{C^2\rho}\right)^{-1}
\sim O(v),
\\
E'&=&(D-1+\Omega L)'\sim O(v),
\\
L'&\sim& O(1),
\\
\Lambda&=&-\frac{C\rho\, u_{\xi} R^2}{4\pi} \sim O(v)\,.
\label{eq:LamNew}
\eeqa 
Then we can show
\beqa
\rho u_{\eta}\left[E'-\Lambda C \Omega'-\Omega L'\right]
=\rho\left[(D-1)'+R v^{\hat \varphi} \Omega'\right].
\label{eq:uoN}
\eeqa
{}From Eqs.~(\ref{eq:LamNew}) and (\ref{eq:LNew}),
we may express $\rho\,u_{\xi}$ as
\beqa
\rho\, u_{\xi}=-\frac{4\pi\Lambda}{CR^2}=
-\frac{4\pi}{CR^2}\frac{L-R\, v^{\hat \varphi}}{C}\,.
\label{eq:ruNew}
\eeqa
Using this expression, we can show
\beqa
\rho u_{\xi}\left[L'-\Lambda C'\right]
&=&-\frac{1}{CR^2}\left(\frac{4\pi L}{C}-R v^{\hat\varphi}\frac{4\pi}{C}\right)
\frac{C}{4\pi}\left[\left(\frac{4\pi L}{C}\right)'
-L\left(\frac{4\pi}{C}\right)'+C\Lambda \left(\frac{4\pi}{C}\right)'\right]
\nonumber\\
&=&-\frac{1}{4\pi R^2}\left(\frac{4\pi L}{C}-
R v^{\hat\varphi} \frac{4\pi}{C}\right)
\left[\left(\frac{4\pi L}{C}\right)'-R v^{\hat \varphi}
\left(\frac{4\pi}{C}\right)'\right],
\label{eq:utN}
\eeqa
where the second equality follows from Eq.~(\ref{eq:LNew}).
It is also easy to show that
 $\rho \Theta u_{\xi}\left[E'-\Lambda(C\Omega)'\right] \sim O(v^3)$.

Therefore from Eqs.~(\ref{eq:J3N}), (\ref{eq:upN}), (\ref{eq:uoN})
and (\ref{eq:utN}), the GS equation in the Newtonian limit
is given by
\beqa
&&\left(1-\frac{4\pi}{C^2 \rho}\right)\Delta^{*}\Psi
-\frac{4\pi}{C}\nabla \left(\frac{1}{C\rho}\right)\cdot
\nabla \Psi
\nonumber\\
&&=
-4\pi\rho R^2\left[(D-1)'+R v^{\hat\varphi} \Omega'\right]
-\left(\frac{4\pi L}{C}-R v^{\hat\varphi} \frac{4\pi}{C}\right)
\left[\left(\frac{4\pi L}{C}\right)'-R v^{\hat\varphi}
\left(\frac{4\pi}{C}\right)'\right]
+4\pi R^2 \rho T S'.
\eeqa
This is equivalent to the Newtonian GS equation in \cite{l86}.

\section{Summary}\label{sec:sum}
We have derived the GS equation (\ref{eq:GS}) 
in noncircular (the most general) stationary axisymmetric spacetimes.
The GS equation has been given in the covariant form projected onto 
the $t=$const. and $\varphi=$const. 2-surface $\Sigma_{t\varphi}$.
We have also derived the wind equation (\ref{eq:uu1})
in noncircular spacetimes.
We have discussed various limits of the GS equation 
(no toroidal field limit, no poloidal field limit,
no magnetic field limit and Newtonian limit).

To obtain equilibrium configurations of magnetars,
we have to solve the GS equation (\ref{eq:GS}).
As first glance, it looks formidable to solve it.
One possibility is to take a perturbative approach to solve the GS equation.
Unless the magnetic field is as strong as the maximum
magnetic field allowed by the virial theorem $\sim 10^{18}$ G
\cite{b95,bg96},
we may assume weak magnetic fields compared with gravity.
Then the magnetic field may be treated as a small perturbation
on an already-known non-magnetized configuration.
This approach is similar to that developed 
for slowly rotating stars \cite{c33,h67},
in which the perturbation parameter is the angular velocity.
Work in this direction is in progress.
The preliminary study indicates that
the degree of noncircularity of the spacetime
in a neutron star with mass $M_{*}$ and radius $R_{*}$,
is about $({N_{\Sigma}}^{\mu} {N_{\Sigma}}_{\mu})^{1/2}
\sim (M_{*}/R_{*}) v_{\rm mf}$
and $({M_{\Sigma}}^{\mu} {M_{\Sigma}}_{\mu})^{1/2}
\sim 0.1$-$0.01 (M_{*}/R_{*}) {\cal R}_{M}$,
where $v_{\rm mf}$ is the velocity of the meridional flow,
${\cal R}_{M}$ is the ratio of the magnetic energy 
to the gravitational energy,
and the length scale is normalized by the mass \cite{Ioka2}.

\begin{acknowledgments}
KI is grateful to F.~Takahara for useful discussions.
This work was supported in part by the
Monbukagaku-sho Grant-in-Aid for Scientific Research,
Nos.~00660 (KI) and 14047214 (MS).
\end{acknowledgments}

\appendix
\section{$(2+1)+1$ decomposition of the energy-momentum tensor}
\label{sec:emt2+1+1}
The $(2+1)+1$ decomposition of the energy-momentum tensor is
\beqa
T^{\mu \nu}
&=&e n^{\mu} n^{\nu}
+j \left(n^{\mu} m^{\nu}+m^{\mu} n^{\nu}\right)
+j^{A} \left(n^{\mu} H_{A}{}^{\nu}+H_{A}{}^{\mu} n^{\nu}\right)
\nonumber\\
&&+s m^{\mu} m^{\nu}
+s^{A} \left(m^{\mu} H_{A}{}^{\nu}+H_{A}{}^{\mu} m^{\nu}\right)
+s^{AB} H_{A}{}^{\mu} H_{B}{}^{\nu}\,,
\label{eq:2+1+1em}
\eeqa
where $n_\mu$ and $m_\mu$ are the unit timelike and spacelike normals
to the 2-surface $\Sigma_{t\varphi}$, respectively, and $A,B=1,2$.
{}For the electromagnetic field in an ideal MHD system,
we have from Eqs.~(\ref{eq:FuB}) and (\ref{eq:ideal}),
$F^{\mu\alpha}F^{\nu}{}_{\alpha}=\left(u^{\mu} u^{\nu}+g^{\mu\nu}\right)B^2
-B^{\mu} B^{\nu}$
and $F^{\mu \nu} F_{\mu \nu}=2B^2$,
and hence
\beqa
\frac{1}{4\pi} \left(F^{\mu \alpha} {F^{\nu}}_{\alpha}
-\frac{1}{4} g^{\mu \nu} F^{\alpha \beta} F_{\alpha \beta}\right)
=\frac{1}{4\pi} \left[\left(u^{\mu} u^{\nu}+\frac{1}{2}g^{\mu\nu}\right)B^2
-B^{\mu} B^{\nu}\right].
\eeqa
Then, using Eqs~(\ref{eq:u2+1+1}) and (\ref{eq:B2+1+1})
the components of the energy-momentum tensor in Eq.~(\ref{eq:2+1+1em})
are obtained as
\beqa
e&=&T^{\mu \nu} n_{\mu} n_{\nu}
=\left(\rho+\rho\epsilon+p\right) \left(u_n\right)^2 - p
+\frac{1}{4\pi}\left[\left\{\left(u_n\right)^2-\frac{1}{2}\right\} B^2
-\left(B_n\right)^2\right],
\label{eq:ee}
\\
j&=&-T^{\mu \nu} n_{\mu} m_{\nu}
=\left(\rho+\rho\epsilon+p\right) u_n u_m
+\frac{1}{4\pi}\left(u_n u_m B^2 - B_n B_m\right),
\label{eq:j}
\\
j^{A}&=&-H^{A}{}_{\mu} T^{\mu \nu} n_{\nu}
=\left(\rho+\rho\epsilon+p\right) u_n u_{\Sigma}{}^{A}
+\frac{1}{4\pi}\left(B^2 u_n u_{\Sigma}{}^{A}-B_n B_{\Sigma}{}^{A}\right),
\label{eq:jA}
\\
s&=&T^{\mu \nu} m_{\mu} m_{\nu}
=\left(\rho+\rho\epsilon+p\right) \left(u_m\right)^2 + p
+\frac{1}{4\pi}\left[\left\{\left(u_m\right)^2+\frac{1}{2}\right\} B^2-
\left(B_m\right)^2\right],
\label{eq:s}
\\
s^{A}&=&H^{A}{}_{\mu} T^{\mu \nu} m_{\nu}=
\left(\rho+\rho\epsilon+p\right) u_m u_{\Sigma}{}^{A}
+\frac{1}{4\pi}\left(B^2 u_m u_{\Sigma}{}^{A}-B_m B_{\Sigma}{}^{A}\right),
\label{eq:sA}
\\
s^{AB}&=&H^{A}{}_{\mu} H^{B}{}_{\nu} T^{\mu \nu}
=\left(\rho+\rho\epsilon+p\right) u_{\Sigma}{}^{A} u_{\Sigma}{}^{B}
+p H^{AB}
+\frac{1}{4\pi}\left[\left(u_{\Sigma}{}^{A} u_{\Sigma}{}^{B}
+\frac{1}{2} H^{AB}\right) B^2 -
B_{\Sigma}{}^{A} B_{\Sigma}{}^{B}\right],
\label{eq:sAB}
\eeqa
where $u_{n}$, $u_{m}$, $u_{\Sigma}{}^{\mu}$,
$B_{n}$, $B_{m}$, $B_{\Sigma}{}^{\mu}$ and $B^2$
are given by Eqs.~(\ref{eq:un}) -- (\ref{eq:uperp}) and
(\ref{eq:Bn}) -- (\ref{eq:BB}),
respectively.

\section{Symbols}\label{sec:symbols}
Here, we summarize definitions of some of the symbols we use, 
which may not be commonly used, with the equation numbers
where they are defined or introduced.

\vspace{3mm}
\noindent
$\bullet$
Quantities conserved along each flow line:
\begin{namelist}{xxxxx}
\item[$\Psi$] Flux function, $\Psi:=A_{\mu}\xi^{\mu}=A_{3}$.

\item[$\Phi$] Electric potential, $\Phi:=A_{\mu} \eta^{\mu}=A_{0}$.

\item[$C$] `Magnetic field strength' relative to the magnitude of
meridional flow;
 Eqs.~(\ref{eq:F31}) -- (\ref{eq:F12}).

\item[$D$] `Fluid energy' per unit mass; Eqs.~(\ref{eq:defD}) and (\ref{eq:DEL}).

\item[$E$] `Energy' per unit mass; Eqs.~(\ref{eq:defE}) and (\ref{eq:defE2}).

\item[$L$] `Angular momentum' per unit mass;
 Eqs.~(\ref{eq:defL}) and (\ref{eq:defL2}).

\item[$\Omega$] `Angular velocity' of the magnetic field line, 
$\Omega=-d\Phi/d\Psi$; Eq.~(\ref{eq:F0A}).

\item[$S$] Entropy per unit mass; Eqs.~(\ref{eq:1law}).

\end{namelist}

\vspace{3mm}
\noindent
$\bullet$
Quantities associated with the metric:
\begin{namelist}{xxxxx}

\item[$\eta^{\mu}$] Killing vector associated with stationarity,
$\eta^\mu=(\partial/\partial t)^\mu$; Eq.~(\ref{eq:eta}).

\item[$\xi^{\mu}$] Killing vector associated with axisymmetry,
$\xi^\mu=(\partial/\partial\varphi)^\mu$; Eq.~(\ref{eq:xi}).

\item[$n^{\mu}$] Unit timelike 4-vector orthogonal to $t=$const. hypersurface
$\Sigma_{t}$; Eqs.~(\ref{eq:defn}).

\item[$h_{\mu\nu}$] 3-metric on $\Sigma_{t}$;
Eqs.~(\ref{eq:defh}).

\item[$m^{\mu}$] Unit spacelike 4-vector orthogonal to $t=$const. and
$\varphi=$const. hypersurface $\Sigma_{t\varphi}$;
Eqs.~(\ref{eq:defm}).

\item[$H_{\mu\nu}$] 2-metric on $\Sigma_{t\varphi}$; Eq.~(\ref{eq:defH}).

\item[$G_{\eta}$] Norm of $\eta^\mu+\Omega\xi^\mu$; Eq.~(\ref{eq:Geta}).

\item[$\Theta$] Quantity such that $\xi^\mu+\Theta\eta^\mu$ is orthogonal
to $\eta^\mu+\Omega\xi^\mu$; Eq.~(\ref{eq:Theta}).

\item[$G_{\xi}$] Norm of $\xi^\mu+\Theta\eta^\mu$; Eq.~(\ref{eq:Gxi}).

\item[$N$] Lapse function, $N=-\eta^\mu n_\mu$; Eqs.~(\ref{eq:defn}).

\item[$N^{\mu}$] Shift vector, $N^\mu=Nn^\mu-\eta^\mu$; Eq.~(\ref{eq:eta}).

\item[$N_{\Sigma}{}^{\mu}$] Projection of $N^{\mu}$ onto 
$\Sigma_{t\varphi}$; Eq.~(\ref{eq:eta}).

\item[$M$] $=m_\mu\xi^\mu$; Eqs.~(\ref{eq:defm}).

\item[$M_{\Sigma}{}^{\mu}$] $=Mm^\mu-\xi^\mu$; Eq.~(\ref{eq:xi}).

\end{namelist}

\vspace{3mm}
\noindent
$\bullet$
Quantities associated with the fluid:
\begin{namelist}{xxxxx}

\item[$u^{\mu}$] Fluid 4-velocity; Eqs.~(\ref{eq:emten}).

\item[$\rho$] Rest mass density; Eqs.~(\ref{eq:emten}).

\item[$\mu$] Enthalpy per unit mass; Eq.~(\ref{eq:defmu}).

\item[$\epsilon$] Internal energy per unit mass; Eq.~(\ref{eq:emten}).

\item[$p$] Pressure; Eq.~(\ref{eq:emten}).

\item[$T$] Temperature; Eq.~(\ref{eq:1law}).

\item[$u_{\eta}$] $=(u^0-\Theta u^3)/(1-\Omega\Theta)$; Eq.~(\ref{eq:uinv}), or
Eqs.~(\ref{eq:u0}) and (\ref{eq:u3}).

\item[$u_{\xi}$] $=(u^3-\Omega u^0)/(1-\Omega\Theta)$; Eq.~(\ref{eq:uinv}),
or Eqs.~(\ref{eq:u0}) and (\ref{eq:u3}).

\item[$u_{n}$] $=-n_\mu u^\mu$; Eqs.~(\ref{eq:u2+1+1}) and
(\ref{eq:un}).

\item[$u_{m}$] $=m_\mu u^\mu$; Eqs.~(\ref{eq:u2+1+1}) and
(\ref{eq:um}).

\item[$u_{\Sigma}{}^{\mu}$] Projection of $u^{\mu}$ onto $\Sigma_{t\varphi}$;
Eqs.~(\ref{eq:u2+1+1}) and (\ref{eq:uperp}).

\item[$\tilde u_{\Sigma}{}^{\mu}$] $u^A$ ($A=1,2$) components of $u^\mu$;
Eq.~(\ref{eq:uinv}).

\item[$\tilde N_{\Sigma}$] A component of $\tilde u_{\Sigma}{}^{\mu}$ 
defined by Eq.~(\ref{eq:tN}).

\item[$\tilde M_{\Sigma}$] A component of $\tilde u_{\Sigma}{}^{\mu}$ 
defined by Eq.~(\ref{eq:tM}).

\end{namelist}

\vspace{3mm}
\noindent
$\bullet$
Quantities associated with the electromagnetic field:
\begin{namelist}{xxxxx}

\item[$E^{\mu}$] Electric field in the fluid rest frame; Eq.~(\ref{eq:E}).

\item[$B^\mu$] Magnetic field in the fluid rest frame; Eq.~(\ref{eq:B}).

\item[$B_{n}$] $=-n_\mu B^\mu=n_\mu u_\nu\epsilon^{\mu\nu\alpha\beta}
F_{\alpha\beta}/2$; Eq.~(\ref{eq:B2+1+1}).

\item[$B_{m}$] $=m_\mu B^\mu=u_\mu m_\nu\epsilon^{\mu\nu\alpha\beta}
F_{\alpha\beta}/2$; Eq.~(\ref{eq:B2+1+1}).

\item[$B_{\Sigma}{}^{\mu}$] Projection of $B^{\mu}$ onto $\Sigma_{t\varphi}$,
Eq.~(\ref{eq:B2+1+1}).

\item[$J^{\mu}$] Electromagnetic current 4-vector; Eq.~(\ref{eq:maxwell2}).

\end{namelist}

\vspace{3mm}
\noindent
$\bullet$
Others:
\begin{namelist}{xxxxx}

\item[$M_{\rm Alf}$] Alfv$\acute{\rm e}$n Mach number; Eq.~(\ref{eq:Malf}).

\item[$\Lambda$] An auxiliary quantity defined by Eq.~(\ref{eq:Lambdadef}).

\end{namelist}








\begin{thebibliography}{123}
\bibitem{k98}
  C.~Kouveliotou et al., Nature (London) {\bf 393}, 235 (1998).

\bibitem{dt92}
  R.~C.~Duncan and C.~Thompson, Astrophys.~J.~Lett. {\bf 392}, L9 (1992).

\bibitem{m02}
  S.~Mereghetti, L.~Chiarlone, G.~L.~Israel, and L.~Stella, astro-ph/0205122.

\bibitem{t01}
  C.~Thompson, astro-ph/0110679.

\bibitem{td93}
  C.~Thompson and R.~C.~Duncan, Astrophys.~J. {\bf 408}, 194 (1993).

\bibitem{u92}
  V.~V.~Usov, Nature (London) {\bf 357}, 472 (1992).

\bibitem{n98}
  T.~Nakamura, Prog.~Theor.~Phys. {\bf 100}, 921 (1998).

\bibitem{kr98}
  W.~Klu$\acute {\rm z}$niak and M.~Ruderman, 
Astrophys.~J.~Lett. {\bf 505}, L113 (1998).

\bibitem{wyhw00}
  J.~C.~Wheeler, I.~Yi, P.~H$\ddot {\rm o}$flich, and L.~Wang,
Astrophys.~J. {\bf 537}, 810 (2000).

\bibitem{b95}
  M.~Bocquet, S.~Bonazzola, E.~Gourgoulhon, and J.~Novak,
Astron.~Astrophys. {\bf 301}, 757 (1995).

\bibitem{bg96}
  S.~Bonazzola and E.~Gourgoulhon,
Astron.~Astrophys. {\bf 312}, 675 (1996).
  
\bibitem{i01}
  K.~Ioka, Mon.~Not.~R.~Astron.~Soc. {\bf 327}, 639 (2001).

\bibitem{kok99}
  K.~Konno, T.~Obata, and Y.~Kojima, 
Astron.~Astrophys. {\bf 352}, 211 (1999).

\bibitem{m99}
  A.~Melatos, Astrophys.~J.~Lett. {\bf 519}, L77 (1999).

\bibitem{c02}
  C.~Cutler, Phys.~Rev.~D {\bf 66}, 084025 (2002).

\bibitem{it00}
  K.~Ioka and K.~Taniguchi, Astrophys.~J. {\bf 537}, 327 (2000).

\bibitem{l67}
  A.~Lichnerowicz, {\it Relativistic Hydrodynamics and Magnetohydrodynamics}
(Benjamin, New York, 1967).

\bibitem{nt73}
  I.~D.~Novikov and K.~S.~Thorne, in {\it Black Holes}, 
edited by C.~DeWitt and B.~S.~DeWitt (Gordon and Breach, New York, 1973).

\bibitem{bo78}
  J.~D.~Bekenstein and E.~Oron, Phys.~Rev.~D {\bf 18}, 1809 (1978).

\bibitem{bo79}
  J.~D.~Bekenstein and E.~Oron, Phys.~Rev.~D {\bf 19}, 2827 (1979).

\bibitem{b93}
  S.~Bonazzola, E.~Gourgoulhon, M.~Salgado, and J.~A.~Marck,
Astron.~Astrophys. {\bf 278}, 421 (1993).

\bibitem{cpl01}
  C.~Y.~Cardall, M.~Prakash, and J.~M.~Lattimer, 
Astrophys.~J. {\bf 554}, 322 (2001).

\bibitem{gb93}
  E.~Gourgoulhon and S.~Bonazzola,
Phys.~Rev.~D {\bf 48}, 2635 (1993).

\bibitem{o02}
  A.~Oron, Phys.~Rev.~D {\bf 66}, 023006 (2002).

\bibitem{p66}
  A.~Papapetrou, Ann.~Inst.~H.~Poincar$\acute{\rm e}$~A {\bf 4}, 83 (1966).

\bibitem{c69}
  B.~Carter, J.~Math.~Phys. {\bf 10}, 70 (1969).

\bibitem{c72}
  B.~Carter, in {\it Black Holes},
edited by C.~DeWitt and B.~S.~DeWitt (Gordon and Breach, New York, 1973).

\bibitem{s03}
  M.~Shibata, gr-qc/0301103.

\bibitem{s00}
  S.~L.~Shapiro, Astrophys.~J. {\bf 544}, 397 (2000).

\bibitem{su00}
  M.~Shibata and K.~Uryu, Phys.~Rev.~D {\bf 61}, 064001 (2000).

\bibitem{meas76}
  D.~L.~Meier, R.~I.~Epstein, W.~D.~Arnett, and D.~N.~Schramm,
Astrophys.~J. {\bf 204}, 869 (1976).

\bibitem{Ioka2}
  K.~Ioka, and M.~Sasaki (in preparation).

\bibitem{m80}
  K.~Maeda, M.~Sasaki, T.~Nakamura, and S.~Miyama,
Prog.~Theor.~Phys. {\bf 63}, 719 (1980).

\bibitem{nok87}
  T.~Nakamura, K.~Oohara, and Y.~Kojima,
Prog.~Theor.~Phys.~Suppl. {\bf 90}, 1 (1987).

\bibitem{s84}
  M.~Sasaki, in {\it Problems of Collapse and Numerical Relativity},
edited by D.~Bancel and M.~Signore (Reidel, Dordrecht, 1984).

\bibitem{l86}
  R.~V.~E.~Lovelace, C.~Mehanian, C.~M.~Mobarry, and M.~E.~Sulkanen,
Astrophys.~J.~Suppl. {\bf 62}, 1 (1986).

\bibitem{ml86}
  C.~M.~Mobarry, and R.~V.~E.~Lovelace,
Astrophys.~J. {\bf 309}, 455 (1986).

\bibitem{n91}
  S.~Nitta, M.~Takahashi, and A.~Tomimatsu,
Phys.~Rev.~D {\bf 44}, 2295 (1991).

\bibitem{b97}
  V.~S.~Beskin, Phys.~Uspekhi {\bf 40}, 659 (1997).

\bibitem{fe37}
  V.~C.~A.~Ferraro, Mon.~Not.~R.~Astron.~Soc. {\bf 97}, 458 (1937).

\bibitem{fe54}
  V.~C.~A.~Ferraro, Astrophys.~J. {\bf 119}, 407 (1954).

\bibitem{wo59} 
  L.~Woltjer, Astrophys.~J. {\bf 130}, 405 (1959).

\bibitem{be85}
  J.~Bekenstein and D.~Eichler, Astrophys.~J. {\bf 298}, 493 (1985).

\bibitem{sy78}
  L.~Smarr and J.~W.~York,
Phys.~Rev.~D {\bf 17}, 2529 (1978).

\bibitem{c86b}
  M.~Camenzind, Astron.~Astrophys. {\bf 162}, 32 (1986).

\bibitem{c86a}
  M.~Camenzind, Astron.~Astrophys. {\bf 156}, 137 (1986).

\bibitem{b70}
  J.~M.~Bardeen, Astrophys.~J. {\bf 162}, 71 (1970).

\bibitem{w72}
  S.~Weinberg, {\it Gravitation and Cosmology} (Wiley, New York, 1972).

\bibitem{c33}
  S.~Chandrasekhar, Mon.~Not.~R.~Astron.~Soc. {\bf 93}, 390 (1933).

\bibitem{h67}
  J.~B.~Hartle, Astrophys.~J. {\bf 150}, 1005 (1967).


\end{thebibliography}
\end{document}